\documentclass[twocolumn]{article}
\usepackage[margin=20truemm]{geometry}
\usepackage[dvips]{graphicx}
\usepackage{latexsym}

\usepackage{listings}
\usepackage{amsmath}
\usepackage{stmaryrd}
\usepackage{color}
\usepackage{url}

\def\Underline{\setbox0\hbox\bgroup\let\\\endUnderline}
\def\endUnderline{\vphantom{y}\egroup\smash{\underline{\box0}}\\}
\def\|{\verb|}

\newcommand{\mifelse}[3]{\mathit{{\rm if}(#1) \; #2 \; {\rm else} \; #3}}

\newcommand{\semantics}[1]{\llbracket #1 \rrbracket}

\newcommand{\mvar}{\mathit{Var}}
\newcommand{\mattr}{\mathit{Attr}}
\newcommand{\msattr}{\mathit{attr}}
\newcommand{\mprim}{\mathit{Prim}}
\newcommand{\mtype}{\mathit{Type}}
\newcommand{\mtypec}{\mathit{Type_c}}

\newcommand{\mmem}{\mathit{Mem}}
\newcommand{\maddr}{\mathit{Addr}}
\newcommand{\mval}{\mathit{Val}}
\newcommand{\mstate}{\mathit{State}}
\newcommand{\mgetmem}{\mathit{getmem}}

\newcommand{\minstance}{\mathit{instance}}

\newcommand{\tajsvr}{${\rm TAJS}_{\rm VR}$}
\newcommand{\query}[1]{\; \langle #1 \rangle \;}

\newcommand{\figref}[1]{\ref{#1}}
\newcommand{\tabref}[1]{\ref{#1}}
\newcommand{\myast}{{\tt\char'052}}

\begin{document}

\title{
  Path-sensitive Type Analysis with Backward Analysis\\
  for Quality Assurance of Dynamic Typed Language Code
}

\author{
Kodama Ryutaro
\url{r.kodama@sde.cs.titech.ac.jp} 
\\
Arahori Yoshitaka
\url{arahori@cs.titech.ac.jp} 
\\
Gondow Kathuhiko
\url{gondow@cs.titech.ac.jp}
}

% \date{$~$Date: 2023/02/12 07:56:28 $\textrm{(+9 for JST)}, ~~$Revision: 1.11 $~$}
\date{}

\maketitle

\begin{abstract}
%動的型つけ言語で書かれたプログラムに対する静的型解析は，正確に速く行うことが難しいという問題がある．
%これは静的に型の情報がないために，変数の近似が難しいからである．
%このため既存の静的型解析器では，正確な解析ができない．

Precise and fast static type analysis for dynamically typed language
is very difficult.
This is mainly because the lack of static type information 
makes it difficult to approximate all possible values of a variable.
Actually, the existing static type analysis methods are imprecise or slow.

%そこで本論文ではPythonで書かれたプログラムへの静的型解析に対して，backward解析を組み合わせることでpath-sensitivityを獲得し，型解析の精度向上を行う．
%これによって，より正確な型情報を入手することができ，静的解析の精度向上が期待できる．\par
%更にこの手法を採用した解析器を実装し，既存の解析器と比較実験を行った．
%すると既存の型解析ツールに比べてFalse Positiveが少なくてより正確な型解析結果を得ることができた．

In this paper, we propose a novel method to improve the precision of 
static type analysis for Python code,
where a backward analysis is used to obtain the path-sensitivity.
%with path-sensitivity obtained by a backward analysis.
By doing so, our method aims to obtain more precise static type information,
which contributes to the overall improvement of static analysis.

To show the effectiveness of our method,
we conducted a preliminary experiment to compare 
our method implementation and the existing analysis tool
with respect to precision and time efficiency.
The result shows our method provides more precise type analysis
with fewer false positives than the existing static type analysis tool.
Also it shows our proposed method increases the analysis time, but
it is still within the range of practical use.
\end{abstract}

%1
%\section{はじめに}
\section{Introduction}
\label{sec:introduction}
%動的型付け言語で書かれたコードに対する静的解析は，precisionとscalabilityの両立が難しいという問題がある．
%これは型の情報がないため，変数の状態の近似が静的型付け言語に比べて難しいからである．
%代表的な先行研究であるMonat et al.\cite{STAofPy}はpath-sensitiveな解析をできない．
%またPyre\cite{Pyre}では，annotationが必要といった制約内でしか，interproceduralかつscalabilityを持った解析はできないといった課題がある．\Par

Static analysis for dynamically typed languages has the problem
that it is difficult to achieve both of precision and scalability.
This is mainly because the lack of static type information makes it 
more difficult to approximate all possible values of a variable
than in statically typed ones.
For example,
Monat et al.,~\cite{STAofPy}, a state-of-the-art analysis,
is not capable of path-sensitive analysis.
Pyre~\cite{Pyre}, another state-of-the-art one, has the problem 
that annotation must be given manually to achieve
inter-procedural and scalable analysis.

%本論文では静的型解析に対して，backward解析を組み合わせることでpath-sensitivityを獲得し，型解析の精度向上を行う．
%backward 解析を採用することで要求駆動で path-sensitive な解析結果を取得できる.
%これによりエラー が発生した箇所等にのみ path-sensitive な解析を
%選択・集中的に行うことができ，効率的に高精度な解析を行うことができる.

In this paper, we propose a novel method to improve the precision 
of static type analysis for Python code,
where a backward analysis is used to obtain the path-sensitivity.
The backward analysis is performed on demand,
so the path-sensitive analysis is performed selectively 
and intensively only where a detailed analysis is needed,
which enables more precise and efficient analysis.

%提案手法は2つのステップからなる．
%1ステップ目ではTAJS\cite{TAJS}を参考にしたflow-sensitiveな静的型解析を前向きに行い，変数が持つ可能性のある型の候補を粗く絞り込む．
%この絞り込みを行うことで，後のbackward解析の回数を減らす効果がある．
%そして次のステップでは，この結果を基に一部の変数に対してbackward解析を行い，その型の候補を実際に取る実行パスが存在するかの確認を行う．
%パスが存在しない場合にはその型の候補を削除する．
%このように実行パスの存在を確認することで，path-sensitivityを獲得することができる．

Our proposed method consists of two steps.
In the first step, a flow-sensitive static type analysis is performed
forwardly (based on the method of TAJS~\cite{tajs09}),
which roughly overapproximates the type candidates of a variable.
This overapproximation has the effect of reducing the number of
the subsequent backward analyses.
The next step performs the backward analysis for a part of variables,
based on the result of the first step, to check whether there exists
an execution path under a type candidate.
When the execution path does not exist, the type candidate is discarded.
Thus, path-sensitivity can be acquired 
by checking the existence (feasibility) of the execution path.

% 本論文では2つの実験を行い，上に述べた手法の有用性を確認した．1つは型解析結果の正確性の確認実験である．この実験では，先行研究のPyreに比べてより正確な型解析結果を得ることができた．もう1つは本手法での解析時間に影響を与える要因の分析実験である．backward解析時のパス数とbackward解析時の1パス当たりのコード行数という2つの要因を設定し，これらを独立に変化させることで，解析時間の変化を調べた．この実験では，backward解析時の1パス当たりのコード行数の方が解析時間に影響を与えるという結果を得た．\par
% 以下に論文の貢献を挙げる．
% \begin{itemize}
%     \item 本論文が扱う動的型付け言語での静的解析を行う上で，既存の問題の背景や先行研究の確認(\ref{sec:background}章)．
%     \item 先行研究の1つであるPyreで解決できない問題に対して，本論文のアプローチの提示(\ref{sec:problem}章, \ref{sec:suggestion}章)．
%     \item 本論文の手法を形式的に定義するための，動的型付け言語のサブセットの定義(\ref{sec:dynamic_language}章)．
%     \item \ref{sec:dynamic_language}章の言語を用いての，forward解析の形式的定義(\ref{sec:forward_formalization}章)．
%     \item \ref{sec:dynamic_language}章の言語を用いての，backward解析の形式的定義(\ref{sec:backward_formalization}章)
%     \item 実装・実験を通して本論文の有用性の確認(\ref{sec:experiments}章)
% \end{itemize}

%本論文では上記提案手法の実装を使って，精度と速度に関して既存研究との比較実験を行い，上に述べた手法の有用性を確認した．
%実験結果では精度が既存研究より高い解析結果を得ることができた．
%一方速度については既存研究よりも増大したが，
%中規模リアル検体でも実用範囲内で解析を行うことができることを確認した．
%以下に論文の貢献を挙げる．
%\begin{itemize}
%    \item path-sensitivityを実現した高精度な型解析手法の提案
%    \item 型解析分野を対象としたbackward解析の形式化
%    \item 提案手法の実装・実験を通して，提案手法の有用性の評価
%\end{itemize}

To show the effectiveness of our method,
we conducted a preliminary experiment to compare 
our method implementation and the existing analysis tool
with respect to precision and time efficiency.
The result shows our method provides more precise type analysis
with fewer false positives than the existing static type analysis tool.
Also it shows our proposed method increases the analysis time, but
it is still within the range of practical use
for real-world medium-sized programs.

The contributions of this paper are as follows:
\begin{itemize}
    \item We propose a novel precise static type analysis method
      with a path-sensitivity.
    \item We formalize the backward analysis for type analysis.
    \item We demonstrate the usefulness of our method
      through our method implementation and the preliminary experiment.
\end{itemize}

%2
%\section{背景}
\section{Background}
\label{sec:background}
% この章では本論文の背景として，静的解析を行う際の課題と，先行研究で解決できていない課題を順に説明する．\par

%\subsection{静的解析}
\subsection{Static Analysis}
\label{subsec:static_analysis}

%静的解析とは，プログラム自体を実行せずに，プログラムの挙動を近似的に把握し，間違った挙動を起こさないか解析を行うことである．
%そのプログラムの挙動の解析のためには，プログラムの途中の状態を近似することが必要不可欠となる．
%なぜならプログラムが取り得る全ての状態を別々に分けると，状態爆発が起こってしまうからである．
%そしてこの近似が正確であるほど，実際のプログラムの挙動に近いものを表現できているので，解析結果はより正確になるが，
%時間資源・空間資源を多く消費するためscalabilityは低くなり，大規模なコードを解析できないという問題がある．
%特に動的型つけ言語コードに対する静的解析では，型解析がとても重要な役割を果たす．
%これは，変数の型情報がある方が，より正確な近似を行いやすいからである．
%このため動的型つけ言語コードの静的解析を行う際には，型解析を行って変数の型を調べることが一般的である．
%しかし既存の手法ではpath-sensitiveな型解析ができず，正確さに欠けるという問題がある．

Static analysis is a technique of analyzing whether a program 
behaves correctly or not without executing the program itself,
by approximating the behavior of the program.
To achieve this,
it is essential to approximate the state of the program in execution.
This is because distinguishing all possible states of a program
separately would result in a state explosion.
And the more precise this approximation is, 
the more precise the results of the analysis will be,
since it represents the behavior closer to the actual one of the program.
However, precise approximation consumes a large amount of time and 
space resources, resulting in low scalability and the inability 
to analyze large codes.
In particular, type analysis plays a very important role 
in static analysis for dynamically typed languages.
This is because it is easier to make more precise approximations 
when the type information of variables is available.
For this reason, when performing static analysis for
dynamically typed language code,
it is common to perform type analysis to determine the type of a variable.
However, the existing methods do not perform path-sensitive type analysis 
and hence they are imprecise.

\subsection{Path Sensitivity}
\label{subsec:path-sensitivity}
%path-sensitivityとは，静的解析の際に，解析結果を実行パス毎に区別することである．
%% 逆に，解析結果を実行パス毎に区別せず，実行パスの合流地点で各実行パスの結果をまとめてしまう解析のことをpath-insensitivityな解析と言う．
%path-sensitiveな解析の方が解析の精度は良くなるが，実行パス毎に状態を分けなくてはならないため，前節で述べたようにscalabilityは低くなるという問題がある．
%このためpath-sensitivityの獲得には，高いscalabilityを両立できる手法が必要となる．

Path-sensitivity in static analysis means that
individual analysis result is computed separately
for each (feasible) execution path.
Although path-sensitive analysis is more precise, 
it requires a separate state for each execution path, 
which reduces its scalability as described in Sec.~\ref{subsec:static_analysis}.
Therefore, to obtain path-sensitivity, 
some technique for scalability is also required.

% \subsection{先行研究}
% \label{subsec:pre_research}
% 動的型つけ言語コードに対する型解析を行うものとしては，Pyre\cite{pyre}やFlow\cite{flow}がある．これらは型推論を用いて変数の型を求める検査ツールであり，特にPyreはPythonを対象にしたものである．ただし，PyreやFlowには解析の精度を上げるための制約が存在する．例えばinterproceduralに解析を行うには，変数や関数の型のannotationが必要となる．また，path-sensitivityも十分にあるとは言えないなどの課題が残っている．この点に関して，\ref{sec:problem}章で具体例を用いて確認する．\par
% 本研究で用いるbackward解析をバグ検知に用いた先行研究として，Thresher\cite{thresher}がある．backward解析とは，プログラム上のある地点から，プログラムを逆向きに実行していき解析を行うことであり，実行パスを1つ1つ見ていくことができるので，path-sensitivityを獲得できる．ThresherではJavaプログラムに対して，Andersenのポインタ解析\cite{pointer-analysis}の精度を向上させるためにbackward解析を用いている．ただしJavaプログラムが対象であるため，本論文で扱っている動的型つけ言語コードに対する型解析の課題を解決できてはいない．\par
% 最後に同じくbackward解析を用いた先行研究として\tajsvr\cite{tajsvr}がある．この研究ではJavaScriptプログラムに対して，静的解析の際にfield-sensitivityを向上させるために，backward解析を用いている．ただしこの研究では，型解析にTAJS\cite{tajs2014}を使っており，path-sensitivityはない．このため，型解析自体の精度向上という課題を解決できてはいない．

%3
%\section{問題設定}
\section{Problem Setting}
\label{sec:problem}
% この章では，既存の型解析研究である\cite{STAofPy}では不正確な解析結果となる例から，本論文で解決する問題について述べる．

To clarify our problem setting,
this section provides our motivating example
(Fig.~\figref{code:motivating_ex}),
for which the existing type analysis~\cite{STAofPy} 
produces incorrect analysis result.

\begin{figure}[tb]
    \begin{lstlisting}
class Create:
  def run(self): ...

class Select:
  def run(self): ...
  def add_where(self): ...

def run_sql(mode):
  # `mode` is `CREATE` or `SELECT`
  if mode == CREATE:
    sql = Create()
  else:
    sql = Select()

  ...

  if mode == SELECT:
    sql.add_where() %\# never called for `Create`%

  sql.run()
    \end{lstlisting}
    \caption{
%      path-sensitivityが十分でない解析により，false positiveが起きてしまう例
      The existing type analysis \cite{STAofPy} produces
      a false positive result 
      ``\texttt{add\_where} can be called even for the instance of \texttt{Create}''
      for this code due to insufficient path-sensitivity.
    }
    \label{code:motivating_ex}
\end{figure}

%\figref{code:motivating_ex}はmypyプロジェクトに存在するコードのうち一部を省略したものであり，Pyreの解析においてpath-sensitivityが十分でないために，false positiveが起きてしまう例である．

% このコードではCreateとSelectの2つのクラスがあり，どちらもrunメソッドが実装してある．
%更にSelectクラスのみadd\_whereメソッドも存在する．そしてrun\_sql関数では引数modeの値に応じてこれらのクラスのインスタンスを作成し，20行目でそのインスタンスのrunメソッドを実行している．
%ただし17行目において，mode変数の値がSELECT，つまりsql変数の型がSelectクラスのインスタンスである場合，add\_whereメソッドを実行する．
%このように17行目のif文の条件により，18行目での変数sqlの型はSelectクラスのインスタンスのみとなることに注意する．
%よってCreateクラスにadd\_whereメソッドは存在しないが，このコードを実行してもエラーが発生することはない．

Fig.~\figref{code:motivating_ex} is an example code,
for which Monat et al.~\cite{STAofPy}  produces a false positive analysis
due to insufficient path-sensitivity.
In this code, there are two classes: \texttt{Create} and \texttt{Select},
both of which have \texttt{run} method.
In addition, only \texttt{Select} class has \texttt{add\_where} method. 
In \texttt{run\_sql} function, instances of these classes are created 
according to the value of the argument \texttt{mode}, 
and the method \texttt{run} of the instance is called at line 20.
However, at line 17,
iff the value of \texttt{mode} is \texttt{SELECT},
that is, iff the type of \texttt{sql} is \texttt{Select class}, 
\texttt{add\_where} method is called.
Note that, according to the condition of the \texttt{if} statement at line 17,
the type of \texttt{sql} at line 18 is always \texttt{Select} class.
Thus, \texttt{add\_where} method
would be never called for the instances of \texttt{Create} class.

%しかしPythonプログラムに対して抽象解釈を行って静的型解析を行う既存研究\cite{STAofPy}では，このコードを正しく解析できず，間違ったエラー(false positive)を出してしまう．
%\cite{STAofPy}はpath-sensitiveな解析をできないため，if文解析後に各パスの状態をマージして解析を続行する．
%よって10$\sim$13行目のif文の解析後，変数sqlの型はCreateまたはSelectクラスのインスタンスであると解析を行う．
%次に17行目のif文では，条件式から変数modeの値を絞り込むことは可能だが，変数sqlの型を絞り込むことはできない．
%よって18行目でも変数sqlの型はCreateまたはSelectクラスのインスタンスであると解析を行ってしまい，Createクラスにはadd\_whereメソッドが存在しないため，attribute errorを出力してしまう．

Unfortunately, a state-of-the-art research~\cite{STAofPy},
which performs static type analysis for Python programs
based on abstract interpretation,
cannot correctly analyze this code and gives a wrong attribute error 
(false positive), that is, ``\texttt{add\_where} can be called even for the instance of \texttt{Create}''.
Because the research~\cite{STAofPy} does not perform path-sensitive analysis, 
it merges the two states of the true/false branch paths 
after analyzing \texttt{if} statements and continues its analysis.
Therefore, after the analysis of the \texttt{if} statement at line 10 to 13,
the variable \texttt{sql} is analyzed as an instance of 
\texttt{Create} or \texttt{Select} class.
Here, the analysis can refine the value of the variable \texttt{mode}
from the conditional expression at line 17,
but it cannot refine the type of the variable \texttt{sql},
since the analysis does not know the fact that the type of \texttt{sql} is \texttt{Select}
iff \texttt{mode} is \texttt{SELECT}.
Therefore, at line 18, the variable \texttt{sql} is analyzed 
as an instance of \texttt{Create} or \texttt{Select} class,
which results in a (wrong) attribute error
because there is no \texttt{add\_where} method in \texttt{Create} class.

%ただし先ほど確認した通り，このエラーはプログラムを実行しても決して発生しない間違ったエラーである．
%以上の例から分かる通り，既存の型解析ツールでは，Pythonプログラムに対してpath-sensitiveな型解析を行うことができず，false positiveを発生させてしまうという問題が存在する．

However, as we have just discussed above, 
this is a wrong error that never occurs when the program is executed.
As we can see from the above example, 
existing type analysis tools are unable to perform path-sensitive 
type analysis for Python programs, resulting in false positives.

%4
%\section{提案手法}
\section{Proposed Method}
\label{sec:suggestion}
%この章では、本論文の提案手法の概要と、前章のコード例(\figref{code:motivating_ex})に対する解析例を説明する。

This section provides an overview of our proposed method,
and an example analysis for the motivating example 
in Fig.~\figref{code:motivating_ex}.

%\subsection{提案手法の概要}
\subsection{Overview of Proposed Method}
\label{subsec:overview}

%本提案手法はforward解析とbackward解析の2つの工程からなる。
%forward解析は変数の型の候補を粗く絞り込む効果があり、backward解析にはpath-sensitivityを獲得する効果がある。
%backward解析を行ってpath-sensitiveな解析を行う先行研究としてThresher\cite{thresher}等が存在するが、Thresherはpointer解析に対してbackward解析を行っており、型解析においてbackward解析を行う点が本提案手法の貢献の1つである。
%ここで``query''を変数と型の関係から生み出される制約条件、``witness''をある初期queryに対し、その状態にまで至る実行パスとして定義する。
%backward解析ではqueryに「そのwitnessを通るために満たされなければならない制約条件」を追加していく。
%そして制約の追加を繰り返す中でqueryがFalseとなった場合（制約に矛盾が表れた場合）、初期queryで表される変数と型の関係を生む実行パスは存在しない、つまり初期queryを満たすwitnessは存在しないという結論を導くことができる。
%このことからpath-sensitiveな型解析結果を取得できる。

Our proposed method consists of two steps:
the forward analysis and the backward analysis.
The forward analysis has the effect that 
it roughly overapproximates down the type candidates of a variable,
while the backward analysis has the effect that
it obtains a path-sensitivity.
Thresher~\cite{thresher} and so on perform the backward analysis 
to obtain path-sensitivity,
but the purpose of Thresher is pointer analysis,  not type analysis.
So one of the contributions of our proposed method is 
that it is the first one, to our knowledge,
that performs the backward analysis for type analysis.

Now we define ``query'' as a constraint created 
by the relationship between variables and types, 
and also defines ``witness'' as an execution path 
that leads to a code location of current interest for a given initial query.
The backward analysis adds constraints to the query,
which must be satisfied to pass through the witness.
Then, when the query becomes unsatisfiable
in the repeated addition of constraints 
(i.e., when a contradiction arises in the constraints), 
we can conclude that there is no execution path that produces 
the relationship between the variable and the type 
represented by the initial query, 
and for this situation, we say
``there is no witness that satisfies the initial query.''
This implies we obtain path-sensitive type analysis result,
since we eliminate infeasible execution paths in the backward analysis.

%\subsection{backward解析の具体例}
\subsection{A Concrete Example of Backward Analysis}
\label{subsec:backward_example}
%\figref{code:motivating_ex}のコード例に対してbackward解析を行う様子を解説する．
%forward解析により18行目の実行直前での変数sqlの型はCreate型またはSelect型という結果を得ているとする．

This section gives a concrete example of backward analysis 
for our motivating example in Fig.~\figref{code:motivating_ex}.
We assume here that, in the forward analysis,
we have already obtained the fact that the type of \texttt{sql} 
at line 18 is \texttt{Create} or \texttt{Select} class.

%\subsubsection{初期queryが「18行目での変数sqlがCreate型」}
\subsubsection{Initial Query is ``the Type of \texttt{sql} at Line 18
is \texttt{Create} Class}
\label{sec:initial query is sql is Create class}

\label{subsubsec:sigs_is_none}
%まずは初期queryを「18行目での変数sqlがCreate型」に設定する．
%実際には18行目で変数sqlがCreate型となる実行パスは存在しないので，このqueryが矛盾となることを導く．

First, we set the initial query to 
``the type of \texttt{sql} is \texttt{Create} class.''
In fact, there is no execution path where the variable \texttt{sql}
becomes of type \texttt{Create} class at line 18,
so the backward analysis reveals that 
this query is \textit{refuted}.

\begin{enumerate}
%    \item 18行目実行直前での初期queryは以下のようになる．
    \item The initial query just before line 18 is:
        \begin{equation*}
            \it{sql} \mapsto \hat{\it{sql}} \land \hat{\it{sql}} == {\rm Create} \notag 
        \end{equation*}

        where $\it{sql} \mapsto \hat{\it{sql}}$ intuitively means
          the abstract value of the variable \texttt{sql} is $\hat{\it{sql}}$,
          and the constraint
            $\hat{\it{sql}} == {\rm Create} \notag$
            must be satisfied.

%    \item 次に17行目のif文の条件式をqueryに追加する．
    \item Then, the conditional expression in the \texttt{if} statement 
      at line 17 is added to the query
      \begin{eqnarray*}
        && \it{sql} \mapsto \hat{\it{sql}} \land \hat{\it{sql}} == {\rm Create} \\
        && \quad \land \; \it{mode} \mapsto \hat{\it{mode}} \land \hat{\it{mode}} == {\rm SELECT}
      \end{eqnarray*}
      % \item 次に10行目でif文の分岐の合流地点に到達する．
      % この場合，then節を通る場合とelse節を通る場合の2つのパスを個別に解析していく．
      % またどちらの場合であっても，節の条件式から得られる制約は，節に入ったときではなく，節から抜けたとき（節内の解析が終わったとき）にqueryに追加する．

    \item Then, the confluence of the \texttt{if} statement branches
      is reached at line 10. 
      The two paths are analyzed separately:
      one through \texttt{then} clause and 
      the other through \texttt{else} clause.
      In either case, the constraint obtained from 
      the conditional expressions in the clause is added to the query 
      when the analysis leaves the clause 
      (i.e., when the analysis in the clause is finished),
      not when the analysis enters the clause.

    \item[3-1] The case through \texttt{then} clause:
      % then節を通る場合
      \begin{enumerate}
        % \item 11行目の代入文からはqueryに変更はない．
      \item The assignment at line 11 does not change the query.
        % \item 10行目の条件式からqueryは以下のようになる．
      \item The conditional expression at line 10 changes the query as follows:
        \begin{eqnarray*}
          && \it{sql} \mapsto \hat{\it{sql}} \land \hat{\it{sql}} == {\rm Create} \\
          && \quad \land \; \it{mode} \mapsto \hat{\it{mode}} \land \hat{\it{mode}} == {\rm SELECT} \\
          && \quad \land \; \hat{\it{mode}} == {\rm CREATE}
        \end{eqnarray*}
        % 変数modeの値がSELECTかつCREATEになることはなく，このqueryは矛盾である．
        % よって「18行目での変数sqlがCreate型」という初期queryを満たす，10$\sim$13行目のif文のthen節を通るwitnessは存在しないことがわかる．
        The value of the variable \texttt{mode} cannot be
        \texttt{SELECT}  and \texttt{CREATE} at the same time,
        and thus this query is refuted.
        Therefore, the analysis knows there is no witness
        of passing through the \texttt{then} clause at line 10 to 13,
        that satisfies the initial query
        ``the type of the variable \texttt{sql} is \texttt{Create} class.''

    \end{enumerate}
    \item[3-2] The case through \texttt{else} clause:
      % else節を通る場合
    \begin{enumerate}
      % \item 13行目の代入文からqueryは以下のようになる．
    \item The assignment at line 10 changes the query as follows:
      \begin{eqnarray*}
        && \it{sql} \mapsto \hat{\it{sql}} \land \hat{\it{sql}} == {\rm Create} \\
        && \quad \land \; \it{mode} \mapsto \hat{\it{mode}} \land \hat{\it{mode}} == {\rm SELECT} \\
        && \quad \land \; \hat{\it{sql}} == {\rm Select}
      \end{eqnarray*}
      % 変数sqlの型がCreate型かつSelect型になることはなく，このqueryは矛盾である．よって「18行目での変数sqlがCreate型」という初期queryを満たす，10$\sim$13行目のif文のelse節を通るwitnessも存在しないことがわかる．
      The type of the variable \texttt{sql} cannot be 
      \texttt{Create} and \texttt{Select} class at the same time,
      and thus this query is refuted.
      Therefore,
      the analysis knows there is no witness of passing through
      the \texttt{else} clause at line 10 to 13,
      that satisfies the initial query ``the type of the variable \texttt{sql}
      is \texttt{Create} class.''
      
    \end{enumerate}
\end{enumerate}

%以上より，18行目実行直前での変数sqlの型からCreate型を取り除くことができ，18行目でattribute errorを出すことがなくなる．
%このように本提案手法ではbackward解析を採用したpath-sensitiveな解析によって，正確な解析を行うことができる．\par

As the result, 
\texttt{Create} can be removed from the type of the variable \texttt{sql}
just before line 18, since the type of the variable \texttt{sql} at line 18
turns out to be only \texttt{Select} class,
and thus the analysis eliminates 
the attribute error at line 18.
So, our proposed method can perform precise analysis 
by employing path-sensitive analysis with the backward analysis. 

%5
%\section{解析の形式定義}
\section{Formal Definition of Analysis}
\label{sec:formalization}

%5.1
%\subsection{forward解析の形式定義}
\subsection{Formal Definition of Forward Analysis}
%forward解析は型の候補を粗く絞るために行う解析であり，flow-sensitiveかつpath-insensitiveに行う．
%これによってbackward解析の実行箇所を絞ることができ，大幅な速度向上を期待できる．

Our forward analysis is flow-sensitive and path-insensitive,
and is used to roughly overapproximate down the type candidates.
This allows the backward analysis to be performed at fewer locations,
which is expected to significantly increase the speed of the analysis.

\begin{figure}[tb]
    \begin{eqnarray*}
        {\rm primitives} && p \in \mprim ::= \mathit{None|True|False}|0 \\
						&& \qquad \qquad \qquad \; |1|...|1.0|...|"{\rm foo}"|... \\
        {\rm variables}  && x, y, z \in \mvar \\
        {\rm attributes} && \msattr \in \mattr \\
        {\rm class \; types} && \tau_c \in \mtypec ::= \mathit{class \; Foo}|... \\
        {\rm types}          && \tau   \in \mtype  ::= {\rm BOOL|INT|FLOAT} \\
            && \qquad \qquad \qquad \; {\rm |STR}|\tau_c \\
		[0.3cm]
        % {\rm memory \; addresses} & \quad & m \in \mmem & = \mvar \cup (\maddr \times \mattr) \\
        % {\rm instance \; addresses} &     & a \in \maddr      & \\
        % {\rm values}                && v \in \mval = \mprim \cup \maddr \cup \mtypec \\
        % {\rm states}                && \sigma \in \mstate = \mmem \hookrightarrow \mval \\
        {\rm abstract \; addresses} && \hat{a} \in \widehat{\maddr} \\
        {\rm abstract \; memories} && \hat{m} \in \widehat{\mmem} = \mvar \cup (\widehat{\maddr} \times \mattr) \\
        {\rm abstract \; values}    && \hat{v} \in \widehat{\mval} = \widehat{\maddr} \cup \mtype \cup \mprim\\
        {\rm abstract \; states}    && \hat{\sigma} \in \widehat{\mstate} = \widehat{\mmem} \hookrightarrow {\cal P}(\widehat{\mval})
    \end{eqnarray*}
	\caption{
          %forward解析でのconcreate domain，abstract domain
          Concrete domain and abstract domain in the forward analysis.}
	\label{fig:forward_domain}
\end{figure}

%提案手法においては型（$\mtype$）をprimitive型，びプログラム中で宣言したクラス（$\mtypec$）と定義する．
%forward解析は型解析を目的としているが，abstract valuesとしてprimitivesも可能となるように定義した．
%これによってより正確な解析が可能となる．
%例えばリストの要素にアクセスする際のindex変数について，primitivesが可能でないと何番目の要素にアクセスしているか分からないが，primitivesが可能であることで何番目の要素にアクセスしているか分かる場合がある．
% つまりelement-sensitiveな解析が可能となる場合があるため，primitivesを可能とすることでより正確な解析を行うことができる．

In our proposed method (Fig.~\ref{fig:forward_domain}) ,
the type ($\mtype$) is defined 
as a primitive type or a class ($\mtypec$) declared in the program.
The forward analysis is intended for type analysis, 
but it is also defined to allow primitives as abstract values,
since this allows for more precise analysis.
For example, suppose that there is a variable \texttt{index} 
to access a list element.
If primitives are not allowed as abstract values,
there is no way to analyze what element of the list is being accessed.
However, if primitives are allowed as abstract values,
there are some cases where the analysis knows it,
e.g., the case there is a conditional \texttt{func(l[index])} in the program.
Thus, primitives as abstract values allow for more precise analysis,
since it increases the cases where the element-sensitive analysis is available.

\begin{figure}[tb]
    \begin{equation*}
		\semantics{.}^{\#} : \; Stmt \rightarrow (\widehat{\mstate} \hookrightarrow \widehat{\mstate})
    \end{equation*}
    \begin{eqnarray*}
      \semantics{x=p}^{\#}(\hat{\sigma})
        & = & \hat{\sigma} [ x \mapsto p ] \\
      \semantics{x=y}^{\#}(\hat{\sigma})
		    & = & \hat{\sigma} [ x \mapsto \hat{\sigma} y ] \\
      \semantics{x=y \oplus z}^{\#}(\hat{\sigma})
        & = & \hat{\sigma} [ x \mapsto \hat{\sigma} y \sqcup \hat{\sigma} z] \\
        % \semantics{x=y.\msattr}^{\#}(\hat{\sigma})  & = & \hat{\sigma} [ x \mapsto 
        %     \{ \hat{v}_{y.\msattr} \in \hat{\sigma} \mgetmem^{\#}(\hat{\sigma} y, \msattr) \\
        % %         & \qquad \qquad \qquad |
        %         &&\hat{a_y} \in \hat{\sigma} y,
        %         \hat{a_y} \in \widehat{\maddr},
        %         \mhasattr^{\#}(\hat{a_y}, \msattr) \}] \\
      \semantics{x=y.\msattr}^{\#}(\hat{\sigma}) && \\
        = \hat{\sigma} [
          x \hspace{-6pt} & \mapsto & \hspace{-6pt} \bigcup_{\hat{a_y} \in \hat{\sigma} y} \hat{\sigma} \mgetmem^{\#}(\hat{a_y}, \msattr)
        ] \\
        \semantics{y.\msattr=x}^{\#}(\hat{\sigma})
			& = & \hat{\sigma} [
				\mgetmem^{\#}(\hat{a_y}, \msattr) \mapsto \hat{\sigma} x
			] \\
			&& \qquad \quad \mathit{for} \; \hat{a_y} \in \hat{\sigma} y \\
		\semantics{x=\mathit{new} \; y()}^{\#}(\hat{\sigma})
			& = & \hat{\sigma} [ x \mapsto \minstance (\hat{\sigma} y) ]\\
		\semantics{s_1;s_2}^{\#}(\hat{\sigma})
		    & = & [\![s_2]\!]([\![s_1]\!](\hat{\sigma})) \\
		\semantics{\mifelse{x}{s_1}{s_2}}^{\#}(\hat{\sigma})
			& = & [\![s_1]\!](\hat{\sigma}) \cup^{\#} [\![s_2]\!](\hat{\sigma})
    \end{eqnarray*}
    \begin{eqnarray*}
        \mgetmem^{\#}  & : & (\widehat{\maddr} \times \mattr) \hookrightarrow \widehat{\mmem} \\
        \minstance     & : & \mtypec \rightarrow \widehat{\maddr} \\
        \sqcup         & : & ( {\cal P}(\widehat{\mval}) \times {\cal P}(\widehat{\mval}) )
			\rightarrow {\cal P}(\widehat{\mval}) \\
        \cup^{\#}      & : & (\widehat{\mstate} \times \widehat{\mstate}) \rightarrow \widehat{\mstate}
    \end{eqnarray*}
    \caption{
      % 抽象解釈に基づくabstract semanticsの定義
    Abstract semantics definition based on abstract interpretation.}
    \label{fig:forward_semantics}
\end{figure}

%次に抽象解釈に基づくabstract semanticsを定義する．
%\figref{fig:forward_semantics}ではPythonの一部のstatementに対して定義を行った．
%リテラル代入（$x=p$）では右辺のprimitivesを，エイリアス（$x=y$）では
%右辺の変数が指しているabstract valuesを，左辺の変数が指すようにする．
%二項演算代入も同様に右辺の演算結果のabstract valueを左辺が指すようにしている．
Now we define the abstract semantics based on abstract interpretation
(Fig~\figref{fig:forward_semantics}).
In Fig~\figref{fig:forward_semantics},
we provide the definitions for typical statements in Python.
For the literal assignments ($x=p$),
the variable $x$ in the left-hand-side (LHS) is defined as
to map to the primitive $p$ in the right-hand-side (RHS).
For the alias assignments ($x=y$),
the variable $x$ in LHS is defined as to map to the abstract values
mapped by the variable $y$.
Similarly, for binary-operator assignments ($x=y \oplus z$),
it is defined as to map to the abstract values calculated in RHS.

%Attribute read（$x=y.attr$）では変数yが指しているオブジェクトのabstract addresses（$\hat{a_y}$）に対し，attr アトリビュートが指すabstract values（$\hat{\sigma} \; \mgetmem^{\#}(\hat{a_y}, \msattr)$）を左辺の変数が指すようにしている．
%逆にattribute writeではattribute readと同様に変数yが指しているオブジェクトのabstract addressesに対し，そのattr アトリビュートが右辺の変数のabstract valuesを指している．
%インスタンス作成（$x=\mathit{new} \; y()$）ではクラスオブジェクトyのインスタンスを左辺の変数が指している．

For attribute read ($x=y.\msattr$),
the variable $x$ in LHS is defined as to map
to abstract values ($\hat{\sigma} \; \mgetmem^{\#}(\hat{a_y}, \msattr)$)
mapped by the attribute $\msattr$ 
for all abstract addresses ($\hat{a_y}$)
of  objects mapped by the variable $y$.
Similarly, for attribute write ($y.\msattr=x$),
the attribute $\msattr$ 
for all abstract addresses of objects mapped by the variable $y$
is defined as to map to abstract values mapped by the variable $x$.
For instance creation ($x=\mathit{new} \; y()$),
the variable $x$ in LHS is defined as to map to the instance
of the class object $y$.

%またシークエンス（$s_1;s_2$）は$s_1$から順に解析を行うことを表し，条件分岐（$\mifelse{x}{s_1}{s_2}$）は各パスの状態をマージすることを表す．
The sequence ($s_1;s_2$) is defined as to analyze in order from
$s_1$ to $s_2$, while
the conditional branch ($\mifelse{x}{s_1}{s_2}$) is defined as to
merge the analysis results of the two branch paths.

%以上のabstract semanticsに従って変数やオブジェクトのアトリビュートのabstract statesを求めていく．
%この形式化によって求められた粗い型解析結果に対し，次節で定義するbackward解析によって型を精錬していく．

In summary,
the above abstract semantics are used to compute the abstract states 
of variables and the attributes of objects 
in the forward analysis.
The rough type analysis results obtained by this formalization 
are then refined by the backward analysis as defined in the next section
(Sec.~\ref{sec:formal definition of backward analysis}).

%5.2
%\subsection{backward解析の形式定義}
\subsection{Formal Definition of Backward Analysis}
\label{sec:formal definition of backward analysis}
%backward解析では考えられる実行パスを1つずつ確認し，query中の制約を更新していって矛盾を導く．
%本節では各statementに対して，その更新のルールをホーア論理によって定義する．

The backward analysis checks each possible execution path 
and updates the constraints in the query to refute the query.
This section defines the rules for the updates for all statements
as Hoare logic.

\begin{figure}[tb]
    \begin{eqnarray*}
        \rm symbolic \; values   && \hat{x}, \hat{y}, \hat{z}, \hat{attr} \in \widehat{Var} \\
        \rm symbolic \; expressions && \hat{e} \in \widehat{Expr} ::= \hat{x}|\hat{v}|p|\hat{e_1} \oplus \hat{e_2} \\
            && \qquad \qquad \qquad \; |\hat{e_1} == \hat{e_2} \\
        \rm heap \; constraints     && H                          ::= \mathit{True}|x \mapsto \hat{x}\\
            && \qquad \; |\hat{y}.attr \mapsto \hat{attr}|H_1 \ast H_2 \\
        \rm pure \; constraints     && P                          ::= \mathit{True}|\hat{e}|P_1 \land P_2| \\
        \rm query                   && Q                          ::= \mathit{False}|H \land P|Q_1 \lor Q_2 \\
    \end{eqnarray*}
    \caption{
      % backward解析のためのドメインの定義
      Defining domains for backward analysis.}
    \label{fig:backward_domain}
\end{figure}

% \figref{fig:backward_domain}にある通りqueryは2種類の制約で表す．
% heap constraintsはプログラム中の変数とシンボリックな変数を結びつける制約である．
% 一方pure constraintsはシンボリック変数間の制約を表すものであり，「ある変数がある型に等しい」という制約や，if文の条件式等から生まれる制約を表すものである．

As shown in Fig.~\figref{fig:backward_domain},
the query is represented by two kind of constraints:
heap constraints and pure constraints.
Heap constraints (e.g., $\it{sql} \mapsto \hat{\it{sql}}$
in Sec.~\ref{subsec:backward_example})
are used to relate a variable in the program
(e.g., \texttt{sql} in Fig.~\figref{code:motivating_ex}),
to a symbolic value (e.g., $\hat{\it{sql}}$
in Sec.~\ref{subsec:backward_example}),
while pure constraints (e.g., $\hat{\it{sql}} == {\rm Create} \notag$
in Sec.~\ref{subsec:backward_example})
are used to represent the constraints on symbolic values
like ``the variable \texttt{sql} must be of type Create.''
Some pure constraints 
(e.g., $\hat{\it{mode}} == {\rm CREATE}$
shown in Sec.~\ref{subsec:backward_example})
come from the analysis of conditionals like \texttt{if} statements.

\begin{figure}[tb]
    \begin{alignat*}{3}
        {\rm Disjunction} & \; \frac{\query{Q_1^{\prime}} s \query{Q_1} \quad
                                     \query{Q_2^{\prime}} s \query{Q_2}}
                                    {\query{Q_1^{\prime} \lor Q_2^{\prime}}
                                     s \query{Q_1 \lor Q_2}} \\[4pt]
        % {\rm Frame} & \; \frac{\query{H_1^{\prime} \land P^{\prime}} s \query{H_1 \land P} \quad
        %                        \mathit{modify}(s, H_2) = \emptyset}
        %                        {\query{H_1^{\prime} \ast H_2 \land P^{\prime}}
        %                        s \query{H_1 \ast H_2 \land P}} \\[4pt]
        {\rm Sequence} & \; \frac{\query{Q_1^{\prime\prime}} s_1 \query{Q_1^{\prime}} \quad
                                 \query{Q_1^{\prime}} s_2 \query{Q_1}}
                                {\query{Q_1^{\prime\prime}} s_1;s_2 \query{Q_1}} \\[4pt]
        {\rm IfElse} & \; \frac{\query{Q_1} s_1 \query{Q} \quad \query{Q_2} s_2 \query{Q}}
                               {\query{(Q_1 \land x) \lor (Q_2 \land !x)}
                                \mifelse{x}{s_1}{s_2} \query{Q}} \\[6pt]
        {\rm Constant} & \; \frac{}{\query{P \land \hat{x}==p}
                                    x=p
                                    \query{P}} \\[6pt]
        {\rm Alias} & \; \frac{}{\query{(y \mapsto \hat{x} \land P ) }
                                 x=y
                                 \query{x \mapsto \hat{x} \land P}} \\[-24pt]
        {\rm Binop} & \;
          \begin{array}{c}
            \\
            \\
            \\
            \hline
            \query{
              y \mapsto \hat{y} \ast z \mapsto \hat{z}
              \land P
              \land \hat{x} == \hat{y} \oplus \hat{z}
            } \\
            x=y \oplus z \\
            \query{x \mapsto \hat{x} \land P}
          \end{array}
        \\[-24pt]
        {\rm AttrRead} & \;
          \begin{array}{c}
            \\
            \\
            \\
            \hline
            \query{y \mapsto \hat{y} \ast \hat{y}.\msattr \mapsto \hat{x} \land P} \\
            x=y.\msattr \\
            \query{x \mapsto \hat{x} \land P}
          \end{array}
        \\[-24pt]
        {\rm AttrWrite} & \;
          \begin{array}{c}
            \\
            \\
            \\
            \hline
            \query{x \mapsto \hat{attr} \land P} \\
            y.\msattr=x \\
            \query{y \mapsto \hat{y} \ast \hat{y}.\msattr \mapsto \hat{\msattr} \land P}
          \end{array}
        \\[-16pt]
        {\rm New} & \;
          \begin{array}{c}
            \\
            \\
            h = y \mapsto \hat{y}
              \quad c = \bigvee_{\tau_c^{\prime} \in \tau_c}(\hat{y}==\tau_c^{\prime}) \\
            \hline
            \query{h \land \hat{\msattr}==\mathit{undef} \land c \land \hat{x} == \hat{y}} \\
            x=\mathit{new} \; y() \\
            \query{h \ast x \mapsto \hat{x} \land P}
          \end{array}
    \end{alignat*}
    \caption{
      % backward解析のための解析ルールの定義
      Defining analysis rule for the backward analysis.}
    \label{fig:backward_rule}
\end{figure}

%\figref{fig:backward_rule}はbackward型解析のための解析のルールを定義したものである．
%Thresher\cite{thresher}と同様にホーア論理を用いて定義を行った．
%ただしbackward解析であるため通常のホーア論理と違い，事後制約から事前制約の方向へ読む必要がある．
%つまり$\query{Q} s \query{Q^{\prime}}$の形で書かれるルールに対し，「statement sに対し，事後制約$Q^{\prime}$が与えられたときに，次の条件を満たす事前制約$Q$を求める．
%$Q$を満たす状態からstatement sを実行した際に，$Q^{\prime}$を満たす状態へと変化する」と解釈を行う．

Fig.~\figref{fig:backward_rule} defines the analysis rules 
for the backward analysis as Hoare logic similar to 
the definitions in Thresher~\cite{thresher}.
However, because this is the backward analysis, 
unlike ordinary Hoare logic, 
it is necessary to read the rules in the direction of
post-constraints to pre-constraints.
So, for a rule of the form $\query{Q} s \query{Q^{\prime}}$,
given a post-constraint $Q^{\prime}$ and a statement $s$,
$Q$ can be deduced from $Q^{\prime}$,
if executing statement $s$ from a state satisfying $Q$ 
produces a state satisfying $Q^{\prime}$.

%\figref{fig:backward_rule}のうち始めの3つはstatementの種類に依存しないルールである．
%ルールDisjunctionは，orで表されるqueryを個々に解析している．
%ルールSequenceではbackward解析でも後の文から1文ずつ解析を行っていけばよいことを表している．
%最後にルールIfElseでは，パスが分岐している場合には個々のパスを1つずつ解析をしている．
%ルールDisjunctionやルールIfElseから分かる通り，backward解析では場合分けが発生する場合に各パスを1つずつ解析するため，これらの分岐を効率的に解析する必要があり，これは今後の課題である．

The first three rules in Fig.~\figref{fig:backward_rule}
are independent of the kind of statements.
The rule Disjunction analyzes each query represented by OR ($\vee$) individually.
The rule Sequence indicates that backward analysis can be performed 
one statement at a time, starting with the last statement.
Finally, the rule IfElse analyzes each path one by one 
when the path branches.
As shown by the rules Disjunction and IfElse, 
the backward analysis analyzes each path one by one 
when a case split occurs, so it is necessary to analyze 
these branches more efficiently, and this is future work.
 
%\figref{fig:backward_rule}の残りのルールは6つは，statementの種類に応じたルールを定義している．
%これら代入のルールに共通していることは，左辺の変数（代入された変数）からsymbolic variableへのマッピング（ex. $x \mapsto \hat{x}$）が事前制約では消されることである．
%その上で，ルールConstantは定数代入であり，symbolic variableが右辺のprimitiveな値と等しいという制約（$\hat{x}==p$）を追加する．
%ルールAliasは変数代入であり，右辺の変数が指すsymbolic variableを，左辺の変数が指していたsymbolic variableに置き換える（$y \mapsto \hat{x}$）．
%ルールBinopは二項演算代入であり，代入後の左辺の変数が指していたsymbolic variableと，代入前の右辺の演算の結果が等しいという制約（$\hat{x} == \hat{y} \oplus \hat{z}$）を追加する．
%ルールAttrReadとルールAttrWriteは共にルールAliasと同様のsymbolic variableの置き換えを行う．
%最後にルールNewでは左辺と右辺の型が等しいという制約（$\hat{x} == \hat{y}$）と共に，attributeがundefであるという制約を追加している．

The remaining six rules in Fig.~\figref{fig:backward_rule}
define the rules corresponding to the kind of statements.
The common point to all these six assignment rules is that the mapping 
from the LHS variable (the assigned variable) to the symbolic value 
(e.g., $x \mapsto \hat{x}$) is erased in the pre-constraint,
since the mapping is the result of the assignment, and thus 
the mapping is useless in the pre-constraint.
The rule Constant for constant assignments
adds the constraint ($\hat{x}==p$)\footnote{
At a glance, this rule seems wrong,
as the value of the variable $x$ before assignment is generally not $p$.
However, this rule is correct and sound technically,
since this rule works to refute the query by adding the constraint $\hat{x}==p$
to the pre-constraint, which actually holds in the post-constraint.
Similar techniques are used in other rules.}
that the symbolic value must be equal to the primitive value on RHS.
The rule Alias for variable assignments ($x=y$)
replaces the symbolic values mapped by the variable on RHS
with the symbolic values ($\hat{x}$) mapped by the variable on LHS
(so, the pre-condition includes $y \mapsto \hat{x}$).
The rule Binop for binary-operator assignments
adds the constraint ($\hat{x} == \hat{y} \oplus \hat{z}$)
that the symbolic value of the variable in LHS after assignment
must be equal to the calculated result of binary-operation in RHS
before assignment.
The rules AttrRead and AttrWrite both replaces
the symbolic value similar to the rule Alias.
Finally, the rule New adds the two constraints.
One is $\hat{x} == \hat{y}$ that the types in LHS and RHS must be equal,
and the other is that the new object's attribute is $\mathit{undef}$
($\hat{\msattr}==\mathit{undef}$).

%6
%\section{実験}
\section{Implementation and Preliminary Experiment}
\label{sec:experiments}

%6.1
%\subsection{実装}
\subsection{Implementation}
\label{subsec:implementation}
%提案手法の実装はJavaを使用してAriadne\footnote{\url{https://wala.github.io/ariadne/}}及びThresher\footnote{\url{https://github.com/cuplv/thresher}}上で行った．
%Ariadneは主にJavaプログラムを対象とした静的解析フレームワークWALA\footnote{\url{https://wala.sourceforge.net/wiki/index.php/Main_Page}}をPython用に拡張したものである．
%ただしAriadneの実装のままでは一部対応できないPythonの文法構造があったため，検体に対して以下のような前処理を行っている．

We implemented our proposed method in Java
based on Ariadne\footnote{\url{https://wala.github.io/ariadne/}}
and
Thresher\footnote{\url{https://github.com/cuplv/thresher}}.
Ariadne is a Python extension of 
WALA\footnote{\url{https://wala.sourceforge.net/wiki/index.php/Main_Page}},
which is a static analysis framework mainly for Java programs.

Note here that there are some Python programs that Ariadne does not support,
so the following preprocessing was performed on the Python programs
when used in the experiment (Sec.~\ref{subsec:experiments}):

\begin{itemize}
%    \item モックモジュールのimport分の追加
  \item Adding \texttt{import} statements for mock modules
%    \item 内包表記をfor文へ変更
  \item Replacing list comprehensions with \texttt{for}-loops
%    \item for-else文をwhile-else文に変更
  \item Replacing \texttt{for-else} statements with \texttt{while-else} statements
%    \item 実引数変数がstar付きであることの明示化

    \item Explicitly specifying with an extra keyword argument
      that an actual argument is with the asterisk (\myast), 
      as Ariadne just ignores the star information.
%      Adding annotations for the arguments with star
%      as  extra keyword arguments
      (e.g., \texttt{func(\myast a)} $\Rightarrow$
      \texttt{func(\myast a, PYPSTA\_STARED\_ARG=a)})
%    \item 一部の累積代入演算子を使うstatementを，算術演算子を使うstatementに変更
    \item Replacing some compound assignment operators
      with not compound assignment operators,
      while preserving the behavior
      (e.g., \texttt{a//=3;} $\Rightarrow$ \texttt{a=a//3;})
\end{itemize}

%本論文で行った形式化はintraproceduralな方法であったが，この実装はinterproceduralな解析ができるように実装している．
%これには以下のように関数呼び出しを扱った．

Although the formalization given in Sec.~\ref{sec:formalization} is
intra-procedural one, our implementation can perform
inter-procedural  analysis.
Our implementation handles the function calls as follows:

\begin{itemize}
    % \item forward解析中での関数呼び出し
    \item Function calls in the forward analysis:
    \begin{itemize}
      % \item 関数呼び出し時には，その関数内部に制御を移して解析を行う．
      %   ただし仮引数は，関数呼び出し時点での実変数の解析結果をあてはめる．
    \item When a function is called, the control of analysis is transferred
      to the inside of the function for further analysis.
      The abstract values of formal parameters
      become the abstract values of actual arguments
      in the analysis results at the time of the function call.
      % \item 関数の最後に到達したら，その関数呼び出し場所に制御を戻す．
      %   返り値がある場合は，その返り値の式の型が関数の返り値の型となる．
      \item When the callee function returns, 
        the control of analysis is transferred 
        to the caller's function call location.
        If there is a return value, 
        the function's return abstract value becomes the abstract value of the return expression.
    \end{itemize}
    %\item backward型解析中での関数呼び出し
    \item Function calls in the backward analysis:
    \begin{itemize}
      % \item 関数呼び出し時には，その関数呼び出し場所をstackに積み，関数内部に制御を移して関数末尾から解析を行う．
      %  （末尾が複数ある場合は，各末尾に対して別々に解析を行う．）
    \item When a function is called, the function call location is pushed
      to the stack, and then the control of analysis is transferred
      to the inside of the function and 
      the analysis is continued from the function tail.
      If there are multiple function tails, 
      the analysis is performed separately for each function tail.
      % \item 関数の最初に到達した場合，以下の場合分けを行う
      \item When the function entry point is reached:
      \begin{enumerate}
        % \item stackが空でない場合，そのstackに積まれている関数呼び出し場所へ戻ってbackward解析を続ける
        \item If the stack is not empty,
          the control of the analysis is transferred to the 
          function call location popped from the stack
          for further backward analysis.
        % \item stackが空の場合，call graphを参照して，その関数を呼び出している場所を全て求め，各呼び出し場所全てについて別々に解析を行う．
        \item If the stack is empty,
          the analysis is performed separately
          for all possible callers obtained from the call graph.
      \end{enumerate}
    \end{itemize}
  \end{itemize}

% 実験ではstackの高さを3までと制限した．これはThresher\cite{thresher}で設けられていたものと同じ高さである．
In the experiment, the stack height was limited to 3. 
This is the same height as that used in Thresher~\cite{thresher}.

%標準ライブラリの解析に対しては，typeshedプロジェクト\footnote{\url{https://github.com/python/typeshed}}を基にsummaryを作成した．
%typeshedプロジェクトはPythonの標準ライブラリとビルトイン関数に型アノテーションをつけたものである．
%これは先行研究のPyre\cite{Pyre}等でも使用されている．
%ただしこのsummaryから得られる抽象状態の粒度は「型」であり，forward解析で定義したabstract valueよりは粒度が粗いため，解析の精度は落ちてしまう可能性がある．
%そこで一部標準ライブラリ及びビルトイン関数については，手書きのsummaryを作成した．\par
%またbackward解析の高速化のため，解析の途中で制約を単純化したり，制約に影響を与えないstatementを飛ばす等の最適化を行っている．

To analyze the programs that use the standard library, 
we created and used \textit{summaries} of the standard library,
based on the typeshed project~\footnote{\url{https://github.com/python/typeshed}}.
The typeshed project is Python standard library and built-in functions 
with type annotations.
This has also been used in previous studies such as Pyre~\cite{Pyre}.
However, the granularity of the abstract states obtained from the summaries
is a ``type'', which is coarser than 
that of the abstract values defined in the forward analysis,
 so the precision of the analysis may be reduced.
To avoid this, we created handwritten summaries for some functions
in the standard library and built-in functions. 

To speed up the backward analysis, some optimizations are performed 
such as simplifying constraints in the middle of the analysis 
and skipping statements that do not affect the constraints.

%6.2
%\subsection{実験}
\subsection{Preliminary Experiment}
\label{subsec:experiments}
%実験に当たって以下の2つのreserch questionを設定した．
%Research questions to investigate in the preliminary experiment are as follows:
Our preliminary experiment aims to answer the following research questions:

\begin{itemize}
%\item RQ1: 提案手法の解析結果は，既存手法と比べてどの程度正確か
\item RQ1: How precise are the analysis results of our proposed method compared to the existing methods?
%\item RQ2: 提案手法の解析時間は，既存手法と比べてどの程度差があるか
\item RQ2: What is the difference in analysis time of our proposed method compared to existing methods?
\end{itemize}

%実験環境は以下の通りである．OSはWindows 10 Home，プロセッサはIntel Core i7-1065G7 CPU@1.30GHz（4コア），RAMは32GBである．\par
%RQ1，RQ2共に比較対象研究として\cite{STAofPy}を採用した．
%実験には合成検体とリアル検体を用意した．
%合成検体はPythonで記述されるコードパターンを念頭に筆者が作成したもの，もしくは\cite{STAofPy}やAriadneで使用されているものである．
%またリアル検体は\cite{STAofPy}で採用されていたリアル検体で実験を行った．
%ただし一部検体では\cite{STAofPy}が対応していないモジュールが含まれており，この場合\cite{STAofPy}の解析対象検体を書き換えて実験を行った．

The environment used in the experiment is: 
Windows 10 Home, Intel Core i7-1065G7 CPU@1.30GHz (4-cores) and  32GB RAM.

Monat et al.,~\cite{STAofPy} 
is employed as a comparison study for both of RQ1 and RQ2.
We used two kinds of benchmarks: synthetic programs and real-world applications.
The synthetic benchmarks are ones that the authors created,
the ones used in Monat et al.,~\cite{STAofPy} and Ariadne.
The real-world benchmarks are the ones used in Monat et al.,~\cite{STAofPy}.
However, some real-world benchmarks in \cite{STAofPy}
were not supported by \cite{STAofPy}.
In such a case, we modified them so as to be analyzed in \cite{STAofPy}.

%6.2.1
%\subsubsection{RQ1: 提案手法の解析結果は，既存手法と比べてどの程度正確か}
\subsubsection{RQ1: How precise are the analysis results of our proposed method compared to the existing methods?}

%本実験では各解析器の解析結果に含まれるfalse positiveの個数を調べる．
%ただし全ての検体が実行時にエラーを出さない検体であるため，静的解析によって検出されたエラーは全てfalse positiveである．
%よって解析で検出されたエラーの個数を調べることとする．
%また\cite{STAofPy}は一部のエラー（KeyError，IndexError，ValueError）について個数をカウントしない実装になっており，正確なエラー個数の比較ができず，またsoundな静的解析を行うには避けられない潜在的エラーが多いため，これらのエラーは計測の対象外とした．\par

This experiment examines the number of false positives 
in each analysis result.
Here, since all benchmarks do not produce errors at runtime, 
all errors detected by static analysis become false positives.
So, we just count the number of errors detected in the analysis.

Also, since some errors (KeyError, IndexError, ValueError) are not counted 
in \cite{STAofPy}, and there are many potential errors 
that cannot be avoided for sound static analysis,
we excluded the errors in the result of the experiment.

\begin{table}[tb] 
    %\caption{精度評価実験結果} 
    \caption{Precision in the experiment}
    \label{tab:rq1_real}
    \hbox to\hsize{
        \hfil
        \begin{tabular}{c|c|cc|c}\hline\hline
                           & LOC & \multicolumn{2}{c|}{our method} & \cite{STAofPy} \\
                           &     &   FP & (refuted) &                          FP \\ \hline
            dict.py*       &   9 &    0 &         1 &                           1 \\
            mutation.py    &  10 &    0 &         0 &                           0 \\
            for.py*        &  10 &    1 &         0 &                           1 \\
            branch.py*     &  12 &    0 &         1 &                           1 \\
            sql.py*        &  29 &	  0 &         1 &                           1 \\
            loop.py*       &  29 &    0 &         1 &                           0 \\
            fannkuch.py    &  54 &    0 &         0 &                           0 \\
            float.py       &  60 & 8(8) &         0 &                        8(8) \\
            coop\_con.py   &  65 &    0 &         0 &                           0 \\
            spectral.py    &  74 &    0 &         0 &                           1 \\
            craft.py       & 132 &    0 &         0 &                           0 \\
            nbody.py       & 156 &    0 &         0 &                           1 \\
            chaos.py       & 309 &    0 &    14(14) &                           0 \\
            richards.py    & 423 & 2(2) &   25(374) &                        2(2) \\
            unpack\_seq.py & 457 &	  0 &         0 &                           0 \\
            \hline
        \end{tabular}
        \hfil
    }
\end{table}

%\tabref{tab:rq1_real}はその実験結果である．
%検体ファイル名に*がついているものは合成検体を表している．
%refutedカラムは，forward解析で発生したfalse positiveのうち，backward解析によって間違ったエラーだと解析できた個数を表している（backward解析は少なくともこの値以上行っている）．
%いくつかの検体で提案手法・\cite{STAofPy}共にfalse positiveを記録しているが，提案手法は\cite{STAofPy}と同等かそれ以下のfalse positiveしか出さないことを確認できた．
%また今回使用した検体はforward解析だけでも精度が高くなる検体が多かったため，backward解析によって型をrefuteできた検体は多くはなかったが，
%それでも合成検体・リアル検体共にbackward解析によって余計なfalse positiveをrefuteし，精度が向上することを確認できた．
Table.~\tabref{tab:rq1_real} shows the result of the experiment,
where program names with an asterisk (\myast) indicate
synthetic benchmarks.
The column ``refuted'' indicates the number of false positives generated 
by the forward analysis that could be analyzed as wrong errors 
by the backward analysis.
Numbers in parentheses indicate the number of false positive
where the same false positive is counted twice for two different call-contexts
(context-sensitive false positives).
In our method, the backward analysis is performed at least these numbers of times.

There are false positives in both of our proposed method and 
Monat et al.,~\cite{STAofPy} method for three programs,
and, for all the three cases,
the number of false positives of our proposed method
are equal to those of Monat et al.,~\cite{STAofPy} method.
For most programs used in the experiment,
using only the forward analysis, not the backward analysis, 
gave sufficiently high precise results.
Because of this,
there are few programs that are refuted by the backward analysis. 
Nevertheless, the result shows that the backward analysis for
both synthetic and real-world programs 
refutes extra false positives and thus 
the backward analysis improves the precision.

%for.py*のfalse positiveはループ内部を必ず実行することが静的に分かるにも関わらず，ループ内部を一度も実行しないパスも解析してしまっているために起きるfalse positiveであった．
%これを正しく解析するには，例えば\texttt{for}文のiteration対象オブジェクト（リスト等）に対し，
%「ループ内部を通ったパス上では要素数が0以上」となるような制約を加えなくてはいけないが，現状のbackward解析ルールでは定義をしておらず，今後の課題である．

The reason for the false positive in for.py* 
is because our proposed method analyzed 
the path that never executes the inside of the loop, 
even though it could be statically known 
that the inside of the loop is always executed at least once.
To correctly analyze this, for example, 
we must add a constraint 
such that ``the number of elements is greater than zero on the path through 
the inside of the loop''
for the objects (lists, etc.) to be iterated in the \texttt{for} statement.
However, this is not defined in the current backward analysis rules 
and this is also future work.

\begin{figure}[tb]
    \begin{lstlisting}
def benchmark(n):
    points = [None] * n
    for i in range(n):
        points[i] = Point(i)
    for p in points:
        p.normalize()
    return maximize(points)
    \end{lstlisting}
    %\caption{提案手法がfalse positiveを出したコード例}
    \caption{Example code where our proposed method gave a false positive}
    \label{code:fp_example}
\end{figure}

%float.pyでも提案手法でfalse positiveが発生した．
%\figref{code:fp_example}は実際のコードの一部である．
%変数\texttt{points}は各要素を\texttt{None}で初期化しているが，
%3$\sim$4行目のループによって全要素が\texttt{Point}クラスのインスタンスとなる．
%しかし現状の解析では，変数\texttt{i}が\texttt{int}型であるということしかわからず，
%リストの全要素が更新されることを解析器は把握できない．
%よって変数\texttt{points}リストは\texttt{None}を要素に含んでいる可能性があると解析を続行してしまい，6行目で\texttt{None}型に対する\texttt{normalize}アトリビュートアクセスがあると解析しfalse positiveを出してしまう．
%変数\texttt{p}が\texttt{None}型とならないことは静的に分かるはずだが，
%変数\texttt{i}及び\texttt{p}に関する制約（例えば「変数\texttt{i}は\texttt{range(n)}の範囲，つまり0$\sim$\texttt{n}-1の値を1つずる取る」という制約など）
%を生み出すbackward解析ルールが不十分であるため，正しい解析ができていない．

Also in float.py, both our proposed method and \cite{STAofPy} gave a false positive.
Fig.~\figref{code:fp_example} is the code snippet from float.py.
Each element of the variable \texttt{points} is initialized with \texttt{None}
at line 2,
and then all elements become instances of the \texttt{Point} class
in the loop at line 3 to 4.
However, our current analysis (and also Monat's analysis)
only knows that the variable \texttt{i} is of type \texttt{int},
and cannot know that all elements of the list are updated.
Thus, the analysis thinks
the list variable \texttt{points} can contain \texttt{None} as an element, 
and it gives a wrong attribute error (false positive)
for the \texttt{normalize} attribute access at line 6,
since \texttt{None} does not have the attribute \texttt{normalize}.
Intuitively,
it is statically known that the variable \texttt{p} cannot be 
of type \texttt{None}, but the analysis does not know this 
due to the lack of some rules in the backward analysis.
For this case, we need the rule that generates the constraint
on the variable \texttt{i} and \texttt{p}
such as ``the variable \texttt{i} has the values of the range
of \texttt{range(n)}, i.e., 
a series of integers from \texttt{0} to \texttt{n-1} with no duplicates.''

%richards.pyのfalse positiveは「全てのパス上で絶対に呼び出されない，とは静的に分からない」例外呼出しである．
%このようなエラーは静的解析である本提案手法の限界とはなってしまう．
%ただし例外呼出しであっても，
%それが静的に絶対に起こり得ないパス上にあるのであれば，
%backward解析によってrefuteすることが可能である．

The one of cause of the false positives in richards.py is an exception.
The analysis cannot statically know that
the exception is never raised on all possible paths.
Such errors are a limitation of our proposed method as a static analysis.
Of course, if the code that can raise an exception is a dead code,
the backward analysis can refute it.

%chaos.pyとrichards.pyではbackward解析による型のrefutationが多く行われた．
%これはエラーが発生したメソッドが多くのコンテキストで呼び出されており，
%その各コンテキスト毎にbackward解析を実行したためである．

In chaos.py and richards.py,
many types are refuted in the backward analysis.
This is because the methods that gave an error are called in many contexts,
and the backward analysis is done for each of those contexts.

%今後の課題としては，現状のbackward解析ルールではコンテナ型アクセスやループ等，
%一般的に静的に解決できない可能性が多い文法についての詳しいルールが定義されていない．
% これらの文法に対してより詳細な制約を生み出すbackward解析ルールの作成が求められる．

The current rules in the backward analysis do not support
some syntax rules, such as the accessing of container data types,
loops, etc, which are generally not statically resolvable.
However, there are some cases where they are statically resolvable
like Fig.~\figref{code:fp_example}.
So we need to refine the rules in the backward analysis
so as to generate more detailed constraints for these syntax rules.
This is future work.

%6.2.2
%\subsubsection{RQ2: 提案手法の解析時間は，既存手法と比べてどの程度差があるか}
\subsubsection{RQ2: What is the Difference in Analysis Time of the Proposed Method Compared to Existing Methods?}

\begin{table}[tb] 
    %\caption{リアル検体に対する解析時間評価実験結果} 
    %\ecaption{Analysis time evaluation experiment results for real specimen}
  \caption{The result of the experiment of analysis time for real-world programs}
    \label{tab:rq2_real}

    \hbox to\hsize{
        \hfil
        \begin{tabular}{c|c|cc}\hline\hline
                           & LOC & our method(s) & \cite{STAofPy}(s) \\ \hline
            mutation.py    &  10 &        1.05  &               0.021 \\
            fannkuch.py    &  54 &        1.25  &               0.077 \\
            float.py       &  60 &        7.92  &               0.083 \\
            coop\_con.py   &  65 &        1.18  &               0.032 \\
            spectral.py    &  74 &        1.37  &               0.22  \\
            craft.py       & 132 &        1.40  &               0.43  \\
            nbody.py       & 156 &        1.50  &               0.028 \\
            chaos.py       & 309 &       12.83  &               2.77  \\
            richards.py    & 423 &      177.72  &               6.25  \\
            unpack\_seq.py & 457 &        9.19  &               5.35  \\
            \hline
        \end{tabular}
        \hfil
    }
\end{table}

%本実験では各解析器の解析時間を3回の実行の平均で比較を行った．
%合成検体はコードサイズが小さいため実験対象から外した．
%結果が\figref{tab:rq2_real}である．
%
%全ての検体で\cite{STAofPy}よりも解析時間は多くなってしまった．
%特にchaos.pyやrichards.pyのようなbackward解析を多く行う検体では，他の検体と比べても突出して多くの解析時間がかかっている．
%ただしこれらのようなbackward解析を多く行う検体でも，実用範囲内の解析時間で解析を行うことを確認できた．
%またAriadneの実装の未成熟さに対する前処理では，
%解析でfalse positiveが発生するような前処理を行わなくてはならない場合がある（ただしbackward解析によってrefuteされる）．
%このためAriadneの実装をより成熟させることで前処理を行わなくて済むようになると，forward解析のfalse positiveが減り，よってbackward解析の回数も減らすことができるので，より高速な解析見込むことができる．

Table.~\figref{tab:rq2_real} shows the result of the experiment of
analysis time, where the average time of three runs for each 
benchmark program is listed.
The programs used in the experiment are all the real-world ones,
since the synthetic ones are all too small.
For all programs, the analysis time of our proposed method is 
more than that of Monat et al., \cite{STAofPy}.
In particular, the backward analysis on programs 
such as richards.py took significantly 
more analysis time than other programs,
since the backward analysis is performed so many times
for richards.py as shown in Table.~\figref{tab:rq1_real}.
However, we consider the analysis time is still within the range of 
practical use 
even for the programs that requires a lot of times of the backward analysis
(177 seconds in Table.~\figref{tab:rq2_real}).

As shown in Sec.~\ref{subsec:implementation},
there are some syntax rules not supported by Ariadne,
so we needed to preprocess (i.e., modify) such Python codes for the experiment,
but we found that the analysis generates false positives
for some preprocessed codes due to this preprocessing
(although they are refuted in the backward analysis).
So the more syntax rules Ariadne supports,
the less preprocessing we need to do,
which leads to even fewer false positives in the forward analysis.
This reduces the number of the backward analysis, 
resulting in faster analysis.

% ただしbackward解析が多いほど解析時間は多くなってしまうことは変わらないため，今後はbackward解析の解析時間の短縮が求められる．例えば各backward解析間で途中の制約の状態等を共有し，パスのマージを行いやすくすることでbackward解析時間を減らすなどの手法が考えられ，今後の課題である．
Of course, a further reduction of the analysis time is necessary,
as a lot of times of the backward analysis is 
the root cause of the analysis time increase.
For example, sharing 
the constraints among different backward analyses
may make it easier to merge multiple path information,
which may result in shorter analysis time.
This is future work.

%7
%\section{参考文献}
\section{Related Work}

\subsection{Type Analysis}
%Pythonプログラムを対象とした静的型解析及び静的型検査を行うものとしては，
%Monat et al.\cite{STAofPy}，Pyre\cite{Pyre}，mypy\cite{mypy}，pytype\cite{pytype}等がある．
%\cite{SAofPy}は\cite{STAofPy}の著者の前論文に当たるインターンシップレポートであり，abstract interpretationの詳しい説明がされている．
%\cite{STAofPy}はPythonのsemanticsに基づいて概ね正確な解析をしているが，\ref{sec:problem}で述べたようにpath-sensitiveな解析ができないという問題がある．
%また\cite{Pyre}はJavaScriptを対象にした静的型解析技術であるFlow\cite{flow}をベースとした解析器であり，一部path-sensitiveな解析もできるが，型アノテーションが必要であったりと十分とは言えない精度の解析結果しか出力できない．
%mypyとpytypeについては\cite{py3wild}で詳しく比較がなされており，特にmypyはアノテーションがないと正確な解析ができず，pytypeはpath-sensitiveな解析ができないため\figref{code:motivating_ex}のコードを正しく解析できないという問題がある．

Monat et al.~\cite{STAofPy}, Pyre~\cite{Pyre}, mypy~\cite{mypy}, 
pytype~\cite{pytype}, etc. are tackling the problem
on static type analysis and static type checking for Python.
The article \cite{SAofPy} is the previous report by the authors
of \cite{STAofPy} and gave a detailed description of the
abstract interpretation.
Monat et al.~\cite{STAofPy} is almost mostly precise based on
Python semantics, but it has the problem that 
it does not perform the path-sensitive analysis as described
in Sec.~\ref{sec:problem}.
Flow~\cite{flow} is a static type analysis method for JavaScript,
and the study~\cite{Pyre} is the type analysis tool for Python
based on Flow~\cite{flow}.
The study~\cite{Pyre} is path-sensitive, but there are some limitations,
for example, manual type annotations are required
and its precision is not enough.
A detailed comparison of mypy and pytype is given in the article \cite{py3wild}.
In particular, mypy cannot correctly analyze the code without annotations, 
and pytype cannot correctly analyze the code 
in Fig.~\figref{code:motivating_ex}
due to the lack of path-sensitivity.

%\subsection{動的型つけ言語コードに対する静的解析}
\subsection{Static Analysis for Dynamically Typed Languages}

%Pythonプログラムを対象にした静的解析器として先ほど挙げた\cite{STAofPy}，
%\cite{SAofPy}の他，
%Fromherz et al.\cite{SVAofPy}，PySA\cite{PySA}，
%PyCG\cite{PyCG}，NoCFG\cite{NoCFG}等がある．
%\cite{SVAofPy}は\cite{STAofPy}の基となる研究であり，TAJS\cite{tajs09}を参考にした抽象解釈によって静的解析を行うが，path-sensitiveな解析はできないという問題がある．
%
%PySAはPyreの技術を基にした静的テイント解析ツールであるが，Pyreと同様に型のアノテーションがないと正しい検知を行うことができなかったり，path-sensitivityがないなどの問題がある．
%またPyCGやNoCFGは，Pythonプログラムに対しての静的call graph解析器である．特にNoCFGは型推論に基づいてcall graphを構築するが，どちらもflow-sensitivityやpath-sensitivityがないなどの問題があり，正確な解析ができない．

In addition to the studies \cite{STAofPy} and \cite{SAofPy},
Fromherz et al.~\cite{SVAofPy}, PySA~\cite{PySA},
PyCG~\cite{PyCG} and NoCFG~\cite{NoCFG} are 
also static analysis tools for Python code.
The study~\cite{STAofPy} is based on \cite{SVAofPy},
which performs the static analysis using the abstract interpretation
based on TAJS~\cite{tajs09}.
However, the study~\cite{SVAofPy} is path-insensitive.
PySA is a static taint analysis tool based on the technique of Pyre,
but it has the problem that 
it cannot correctly analyze the code without annotations as well as Pyre, 
and it is path-insensitive.
PyCG and NoCFG are static call graph generators for Python code.
NoCFG utilizes the type inference in the call graph generation.
Both analyses in PyCG and NoCFG are flow-insensitive and path-insensitive,
and thus their analyses are imprecise.

% JavaScriptを対象にした静的解析器ではTAJS\cite{tajs2014,tajs09}や\tajsvr \cite{tajsvr}がある．
%TAJSはJavaScriptで静的型解析を行う際のフレームワークとして多く用いられているが，このうちTAJS\cite{tajs09}はscalabilityが低く，path-sensitivityがないなどの問題がある．
%\cite{tajs2014}ではparameter sensitivityやloop specializationによって，変数のabstract valueをより細かく解析する方法が取られており，本研究のforward解析に適用することでより正確な解析ができる可能性がある．
%また\tajsvr はこのTAJSを使って，JavaScriptオブジェクトに対してbackward解析によるfield-sensitivityの向上を行ったものである．
%ただしTAJSを使っているため型解析の結果はpath-insensitiveである．一方，本手法を使って静的解析器を作成した場合，path-sensitiveな型解析結果を得ることができるため，その静的解析の結果自体の精度向上を見込むことができる．

TAJS~\cite{tajs2014,tajs09} and \tajsvr~\cite{tajsvr}
are static type analysis tools for JavaScript code.
TAJS is widely used as a framework of the static type analysis for JavaScript,
but it has the problem that it is not scalable and path-insensitive.
The study \cite{tajs2014} uses the techniques
like the parameter-sensitivity and the loop specialization
to more precisely analyze the abstract values of the variables.
These techniques can be applied to the forward analysis of our proposed method
to improve the precision.
\tajsvr is based on TAJS, and improves the field-sensitivity
for JavaScript objects by the backward analysis.
However, the result of type analysis is path-insensitive,
since it is based on TAJS.
This problem may be resolved by the path-sensitive type analysis 
in our proposed method.

% JavaScriptを対象にした静的Call Graph解析器の研究としては，ACG\cite{ACG}やreserch paperである"Static JavaScript Call Graphs"\cite{static_JACG}がある．\cite{static_JACG}ではACGやTAJSも含めた5つの解析器の比較を行っており，型解析のアプローチは精度が良い手法であることが示されている．しかしこの論文で挙げられている5つの手法のどれもpath-sensitiveな解析は出来ないため，精度の面で未だ改善点がある．\par
% 動的型つけ言語であるJavaScriptのプログラムに対しての静的解析の困難さを述べているものにAnalysis of JavaScript Programs\cite{ana-js}がある．この文章は近年のJavaScriptの研究の傾向についてまとめられており，その中で静的解析の際に障壁となる事柄についてもまとめられている．その内容は本手法の対象であるPythonにも通じるものがあり，背景知識として役に立つものである．ただしこの文章は調査分析のみで，新しい手法の提案までは行っていない．

%\subsection{backward解析を採用した静的解析}
\subsection{Static Analysis using Backward Analysis}

%本研究のようにbackward解析によって精度の向上を行った研究としてThresher\cite{thresher}，\tajsvr \cite{tajsvr}，SUPA\cite{SUPA}等がある．
%Thresherは本提案手法と同様にホーア論理に基づく形式手法によって，ポインタ解析の結果の精錬を行っている．
%一方SUPAはflow-sensitiveなポインタ解析を高速に行うことを目標としており，こちらもポインタ関係の精錬のためにbackward解析を採用している．
%どちらも静的型付け言語を対象としたポインタ解析での利用であり，本研究の動的型付け言語を対象とした型解析とは異なるものである．

Like our proposed method,
Thresher~\cite{thresher}, \tajsvr~\cite{tajsvr} and SUPA~\cite{SUPA}
use the backward analysis to improve the precision.
Thresher refines the result of pointer analysis
by the formalization based on Hoare logic 
(our proposed method also uses a similar formalization based on Hoare logic).
SUPA aims to perform fast flow-sensitive pointer analysis,
and also uses the backward analysis to refine the pointer information.
They are the same as our proposed method in that
they use the backward analysis,
but differ in their intended purposes:
the pointer analysis in Thresher and SUPA,
and the type analysis for dynamically typed languages in our proposed method.

%8
%\section{おわりに}
\section{Conclusion}

%本研究は動的型付け言語を対象とした静的型解析の精度向上のため，backward解析を採用してpath-sensitiveな解析を行う手法を提案した．
%backward解析では要求駆動な形でパスを1つずつ解析するため，高精度（path-sensitive）な解析を効率的に行うことができる．
%実験では既存手法との比較を行い，精度は向上していることを確認できた．
%解析時間については既存手法と比べて増加したが，実用範囲内の解析時間で解析できることも確認できた．

This paper proposed a novel method to improve the precision of 
static type analysis for dynamically typed languages.
To improve precision,
our proposed method performs a path-sensitive analysis with the backward
analysis.
The backward analysis analyzes the paths one at a time on demand,
which enables fast path-sensitive analysis.

The preliminary experiment shows 
our proposed method improved the precision, 
compared to the existing static type analysis tool.
Also it shows our proposed method increases the analysis time, but
it is still within the range of practical use.

%今後の課題としては，コンテナ型やループに関するより詳細なbackward解析ルールを定義して精度の更なる向上を目指すことや，backward解析間の状態の共有によりbackward解析時間の短縮等が挙げられる．

Our future work includes
even higher precision by defining the detailed rules
in the backward analysis for container data types and loops,
and even shorter analysis time by sharing the constraints
among different backward-analyses.

% \bibliography{bibsample}

\begin{thebibliography}{10}

\bibitem{STAofPy}
Monat, Rapha{\"e}l and Ouadjaout, Abdelraouf and Min{\'e}, Antoine:
Static Type Analysis by Abstract Interpretation of Python Programs.
In 34th European Conference on Object-Oriented Programming (ECOOP),
Vol. 166 of Leibniz International Proceedings in Informatics (LIPIcs),
Berlin (Virtual / Covid), Germany, November 2020.
Schloss Dagstuhl--Leibniz-Zentrum f{\"u}r Informatik.

\bibitem{Pyre}
Meta:
Pyre, a performant type-checker for Python 3. (online),
\url{https://pyre-check.org/}
(2023.02.06).

\bibitem{tajs09}
Jensen, Simon Holm and M\o{}ller, Anders and Thiemann, Peter:
Type Analysis for JavaScript.
Proceedings of the 16th International Symposium on Static Analysis (SAS),
August, 2009.
Springer-Verlag.

\bibitem{thresher}
Blackshear, Sam and Chang, Bor-Yuh Evan and Sridharan, Manu:
Thresher: Precise Refutations for Heap Reachability.
SIGPLAN Not., Vol.48,
New York, NY, USA, June 2013.
Association for Computing Machinery.

%国際会議なので，国内会議の参考文献は削除
%\bibitem{pypstaCG}
%児玉龍太郎, 荒堀喜貴, 権藤克彦:
%動的型付け言語を対象とする正確な型解析に基づく手続き間制御フロー解析とその応用.
%第25回プログラミングおよびプログラミング言語ワークショップ (PPL) 2023

\bibitem{mypy}
the mypy project:
"Mypy: Static Typing for Python (online),
\url{https://www.mypy-lang.org/}
(2023.02.06).

\bibitem{pytype}
google:
pytype (online),
\url{https://google.github.io/pytype/}
(2023.02.06).

\bibitem{SAofPy}
Monat, Rapha{\"e}l:
Static Analysis by Abstract Interpretation Collecting Types of Python Programs.
Internship report,
LIP6 - Laboratoire d'Informatique de Paris 6,
September 2018.

\bibitem{flow}
Chaudhuri, Avik and Vekris, Panagiotis and Goldman, Sam and Roch, Marshall and Levi:
Fast and Precise Type Checking for JavaScript.
Proc. ACM Program. Lang., Vol.1, No.OOPSLA, October 2017.
Association for Computing Machinery.

\bibitem{py3wild}
ak-amnouykit, Ingkarat and McCrevan, Daniel and Milanova, Ana and Hirzel, Martin and Dolby, Julian:
Python 3 Types in the Wild: A Tale of Two Type Systems.
Proceedings of the 16th ACM SIGPLAN International Symposium on Dynamic Languages,
New York, NY, USA, 2020.
Association for Computing Machinery.

\bibitem{SVAofPy}
Fromherz, Aymeric and Ouadjaout, Abdelraouf and Min{\'e}, Antoine:
Static Value Analysis of Python Programs by Abstract Interpretation.
In NFM 2018 - 10th International Symposium NASA Formal Methods,
Vol. 10811 of Lecture Notes in Computer Science,
pp.185–202, Newport News, VA, United States, April 2018.
Springer.

\bibitem{PySA}
Meta:
PySA Overview (online),
\url{https://pyre-check.org/docs/pysa-basics/}
(2023.02.06).

\bibitem{PyCG}
Salis, Vitalis and Sotiropoulos, Thodoris and Louridas, Panos and Spinellis, Diomidis and Mitropoulos, Dimitris:
PyCG: Practical Call Graph Generation in Python.
In Proceedings of the 43rd International Conference on Software Engineering,
ICSE ’21, pp.1646–1657.
IEEE Press, 2021.

\bibitem{NoCFG}
Abadi, Aharon and Makovitzki, Bar and Shemer, Ron and Tyszberowicz, Shmuel:
A Lightweight Approach for Sound Call Graph Approximation.
In Proceedings of the 37th ACM/SIGAPP Symposium on Applied Computing, SAC ’22,
p. 1837–1844, New York, NY, USA, 2022.
Association for Computing Machinery.

\bibitem{tajs2014}
Andreasen, Esben and Mller, Anders:
Determinacy in Static Analysis for JQuery.
SIGPLAN Not., Vol.49, No.10, October 2014.

\bibitem{tajsvr}
Stein, Benno and Nielsen, Benjamin Barslev and Chang, Bor-Yuh Evan and M\o{}ller, Anders:
Static Analysis with Demand-Driven Value Refinement.
Proc. ACM Program. Lang., Vol.3, No.OOPSLA, October 2019.

\bibitem{SUPA}
Sui, Yulei and Xue, Jingling:
On-Demand Strong Update Analysis via Value-Flow Refinement.
In Proceedings of the 2016 24th ACM SIGSOFT International Symposium on Foundations of Software Engineering,
FSE 2016, p.460–473, New York, NY, USA, 2016.
Association for Computing Machinery

\end{thebibliography}
% \bibliographystyle{ipsjunsrt}

\end{document}